\documentclass[journal=jpca,manuscript=article]{achemso}
\SectionNumbersOn
\setkeys{acs}{articletitle = true}
\usepackage{graphicx}
\usepackage{float}
\usepackage{bm}
\usepackage{siunitx}
\usepackage{color}
\usepackage{ulem}
\usepackage{tabularx}
\usepackage{multirow}
\usepackage{longtable}
\usepackage{adjustbox}
\usepackage{threeparttable}
\usepackage{blindtext}
\usepackage{comment}

\makeatletter 
\newcommand\LastLTentrywidth{1em} 
\usepackage[version=3]{mhchem} 
\usepackage[usenames,dvipsnames]{xcolor}
 \usepackage{amsmath}
\usepackage{relsize}
\usepackage{mathabx}

\newlength\longtablewidth 
\newcommand\getlongtablewidth{%
	\begingroup 
	\ifcsname LT@\roman{LT@tables}\endcsname 
	\global\longtablewidth=0pt 
	\renewcommand\LT@entry[2]{\global\advance\longtablewidth by ##2\relax\gdef\LastLTentrywidth{##2}}%
	\@nameuse{LT@\roman{LT@tables}}%
	\fi 
	\endgroup} 
\makeatother

\author{Sandra~G\'omez}

\affiliation{Department of Chemistry, University College London, 20 Gordon Street, London WC1H 0AJ, United Kingdom}
\alsoaffiliation{Institute of Theoretical Chemistry, Faculty of Chemistry, University of Vienna, Währingerstrasse 17, 1090 Vienna, Austria}
\alsoaffiliation{Current address: Departamento de Qu\'imica F\'isica, University of Salamanca, 37008, Spain }

\email{sandra.gomez@usal.es 
}

\author{Nadja~K.~Singer}
\affiliation{Institute of Theoretical Chemistry, Faculty of Chemistry, University of Vienna, Währingerstrasse 17, 1090 Vienna, Austria}

\author{Leticia~Gonz\'alez}

\affiliation{Institute of Theoretical Chemistry, Faculty of Chemistry,  University of Vienna, Währingerstrasse 17, 1090 Vienna, Austria}

\author{Graham~A.~Worth}

\affiliation{Department of Chemistry, University College London, 20 Gordon Street, London WC1H 0AJ, United Kingdom}

\title{Do we need delocalised wavefunctions for the excited state dynamics of 1,1-difluoroethylene?}

\begin{document}





\pagebreak
\begin{abstract}
In this work we set up a model Hamiltonian to study the excited state quantum dynamics of 
1,1-difluoroethylene, a molecule that has equivalent atoms exchanged by a torsional symmetry operation leading to equivalent minima on the potential energy surface. In systems with many degrees of freedom where the minimum energy geometry is not unique, the ground state wavefunction will be delocalised among multiple minima. In  this small test system, we probe the excited state dynamics considering localised (in a single minimum) and delocalised (spread over among multiple minima) wavefunctions and check whether this choice would influence the final outcome of the quantum dynamics calculations. Our molecular Hamiltonian comprises seven electronic states, including valence and Rydberg states, computed with the MS-CASPT2 method and projected onto the vibrational coordinates of the twelve normal modes of 1,1-difluoroethylene in its vibrational ground state. This Hamiltonian has been symmetrised along the torsional degree of freedom to make both minima completely equivalent and the model is supported by the excellent agreement with the experimental absorption spectrum. Quantum dynamics results show that the different initial conditions studied do not appreciably affect the excited state populations or the absorption spectrum when the dynamics is simulated assuming a delta pulse excitation. 

\end{abstract}
\textbf{keywords:} quantum dynamics, excited states, 1,1-difluoroethylene, superpositions


\section{Introduction}

Symmetry is a powerful tool used to simplify problems and predict properties in a wide range of physical systems. In the context of quantum mechanics applied to molecules, symmetry operations must leave the molecular Hamiltonian invariant, i.e., it always belongs to the totally symmetric irreducible representation. For small molecules belonging to non-Abelian symmetry groups, the probability of finding the system in a particular conformation must be exactly the same as finding it in a conformation equivalent by symmetry. The nuclear wavefunction and potential energy surfaces must then be constructed taking the permutations of the symmetry group into account. The global delocalised wavefunction can be expressed as a superposition of localised wavefunctions at the single minima transformed via symmetry operations. These superpositions have been detected in the laboratory since the Young double slit experiment\cite{young1802} and much later in buckyballs,~\cite{Arndt1999} and in systems up to 800 atoms.~\cite{Eibenberger2013}. 
As soon as the system moves away from this relaxed state (i.e., stops being an eigenstate), the expectation values of observables and the wavefunction itself change with time. A question would be whether parts of the superposition might interact differently, maybe due to their relative phase sign, when compared to wavefunctions that are localised in a single minima.

One interesting application where nuclear superpositions are invoked is the theoretical description of nuclear spin isomers (NSIs).
The 1920s brought NSIs into the limelight, when the theory about the existence of two different species of the hydrogen molecule (ortho and para) was developed.~\cite{hund27,heis27} However, even now the separation, identification, and conversion mechanisms between NSIs is a challenging problem and only a few polyatomic molecules (CH$_3$F \cite{Panfilov1983,Chapovsky1985,Bakarev1986,Chapovsky1990}, $^{13}$CH$_3$F \cite{13ch3fnagels,Nagels1997,Cacciani2004}, H$_2$CO \cite{H2CO}, $^{13}$CCH$_4$ \cite{13CCH4}, CH$_3$OH \cite{ch3OH}, H$_2$O, \cite{Konyukhov1986,Tikhonov2002,H2O,Horke2014,Kilaj2018} and C$_2$H$_4$ \cite{C2H4,C2H42}) have been separated in their isomeric forms. 
In the last ten years theories connecting NSIs with superpositions of wavepackets have been developed.~\cite{Grohmann2011,Flos2012,Obaid2014,Obaid2015,Belz2013,Grohmann2018,Dongming2018, Xiao-Quin2021,Yang2022} 
Using reduced dimensionality models, an association between the nuclear spin and an intermolecular torsion has been made\cite{gomez2017,Deeb2007,Grohmann2007,Belz2009,Belz2013,Grohmann2011,Flos2012} and  interference of torsional wavepackets within these reduced models leads to different excited state dynamics that can be associated to one or the other NSI, allowing their differentiation.\cite{Obaid2014,Obaid2015,Waldl2016}

Even though the connection between nuclear spin isomers and rotational levels has been extensively reported, the relation to torsional superpositions (as part of the vibrational wavefunction) has been mostly studied from the theoretical point of view. In this work we will focus on torsional superpositions as initial conditions for the dynamics, without making further connections to NSIs. 
To the best of our knowledge, every study that has been done using torsional superpositions in the excited state has used reduced dimensionality models. This is the first time that a full dimensional system is considered, using the case of the 1,1-difluoroethylene (1,1-DFE) molecule in the gas phase, which possesses twelve internal degrees of freedom. 

In excited state dynamics, the choice of the initial conditions can critically affect the outcome of a simulation.\cite{Barbatti2015,Suchan2018,Avagliano2022} For mixed quantum-classical trajectory methods, such as Tully surface hopping\cite{Tully1971} or ab initio multiple spawning\cite{fmsbook2002}, in gas phase simulations geometries and velocities  are usually selected from a Wigner distribution that mimics the lowest eigenstate of the ground state quantum harmonic oscillator. In direct dynamics variational multiconfigurational Gaussian (DD-vMCG)\cite{worth2015} methods, a single multi-dimensional Gaussian function that is the solution for the ground state quantum harmonic oscillator carries the initial population. 
Grid methods like multiconfigurational time dependent Hartree (MCTDH)\cite{Beck1997a} are very flexible and almost any type of initial wavefunction can be formed and used in the propagation.

For this reason, here dynamics simulations were carried out using the MCTDH method,~\cite{Beck2001,wor08:569} treating explicitly all degrees of freedom at a quantum level. Three different initial conditions were considered: two superpositions, one with a positive relative phase and one negative, and a wavefunction localised in a single minimum, i.e., without applying any symmetry operations on it. The deactivation of the population from the valence $\pi\pi*$ state towards the Rydberg state is found to be not particularly affected by the choice of the initial conditions, i.e., superpositions give the same outcome as starting the dynamics in a localised minima. If this behaviour can be extrapolated to larger systems with multiple minima, this result implies an enormous saving in computer time, since only a small part of the configuration space needs to be initially covered. 

To describe the full-dimensional potential energy surfaces (PES) of  1,1-DFE, we use a vibronic coupling model~\cite{Koeppel} up to sixth order and fitted to high level ab-initio quantum chemistry calculations at different geometries. Since torsional angles of 90 degrees are needed to arrive at the conical intersections present in this system, the corresponding normal mode needs to be transformed to the actual dihedral angle, and high-order coupling terms (up to the sixth order) need to be included to describe the processes involving this motion with accuracy. 
In an earlier study on 1,1-DFE using multistate complete active space perturbation theory up to second order (MS-CASPT2) in MOLCAS, we found that a number of Rydberg states must be included to correctly describe the non-adiabatic dynamics starting from the bright valence state.\cite{gomez2019} 

To prove that our Hamiltonian is indeed a good model, we calculate the absorption spectrum after excitation to the excited-state manifold. The excellent agreement between theoretical and experimental spectrum supports our choice of a Hamiltonian and allows the assignment of the spectral bands. Using a localised initial condition and starting the excited state dynamics from the zwitterionic $\pi*^2$ and Rydberg $\pi$-3s, $\pi$-3p$_x$, $\pi$-3p$_y$, and $\pi$-3p$_z$ states,  a reduced dimensionality model then allows us to theoretically unravel the origin of the spectral progression of bands detected in the experimental absorption spectrum of 1,1-DFE.\cite{limao2006} 

\section{Methods}

\subsection{Construction of the nuclear Hamiltonian}
   
To analytically describe the PES, we used a vibronic coupling Hamiltonian \cite{Koeppel,wor04:127}. 
This method, diabatic by ansatz, expresses the PES as a Taylor expansion around a given point, normally the optimised geometry (Frank-Condon point). The Hamiltonian $\bm{H}$ can be thus defined as
\begin{align}
	&\bm{H} = H^{(0)} \bm{1} + \bm{W}^{(0)} + \bm{W}^{(1)} + \bm{W}^{(2)} + \bm{W}^{(3)} + ... ,
\label{eq:1}
\end{align}
where each matrix has a dimension of 7, the number of electronic states involved. As a coordinate basis to express the Hamiltonian, we choose the mass-frequency scaled ground-state normal vibrations, which in the case of 1,1-DFE means twelve normal modes. The zero-order Hamiltonian $H^{(0)}$ corresponds to the kinetic energy operator. For the normal modes q$_1$-q$_3$ and q$_5$-q$_{12}$, the kinetic energy operator $\hat{T}_q$ reads 
\begin{equation}
	\hat{T}_q = -\sum^{12}_{\alpha=1}\hspace*{-0.1cm}{}^\prime \, \frac{\omega_{\alpha}}{2}  \frac{\partial^2}{\partial Q^2_{\alpha}},
\label{eq:2}
\end{equation}
where $\omega_\alpha$ is the ground-state normal mode
frequency and the prime on the summation indicates $\alpha=4$ is omitted.

The normal mode q$_4$ is replaced by the torsion angle $\theta$ between the CH$_2$ group and the molecular plane, considering the fluoride atoms to be fixed due to the large difference in atomic masses. Since a torsional degree of freedom implies a complete rotation along the symmetry axis whereas the normal mode is a harmonic back and forth motion, we transformed the kinetic energy operator following reference \citenum{cat01:2088}. The kinetic energy operator $\hat{T}$ for this degree of freedom now reads
\begin{equation}
 \hat{T}_{\theta} = -\frac{1}{2\mu r^2}\frac{\partial^2}{\partial\theta^2},
 \label{eq1}
\end{equation}
where $\mu r^2$ corresponds to the moment of inertia with a mass $\mu = 2$\,amu, and $r$ is the ground-state equilibrium distance from the hydrogen atoms to the inertial axis (along the C=C bond), which is $0.974$\,\AA. The total zero-order Hamiltonian, $H^{(0)}$, is then described as
\begin{equation}
 H^{(0)}(Q) = \hat{T}_{\theta} + \hat{T}_q, 
\end{equation}
the sum of both terms.
 
The matrices $\mathbf{W}$ form the diabatic potentials and couplings. Following the standard vibronic coupling model scheme, for modes that are close to harmonic (q$_1$, q$_3$, q$_5$, q$_7$, q$_8$, q$_9$ and q$_{12}$), the zeroth order potential energy matrix $W^{(0)}_{ij}$ is diagonal and formed by the ground-state harmonic oscillator potential shifted by the vertical energies:
\begin{alignat}{2}
	&W^{(0)}_{ij}(Q) &&= \left( E_i + V \right) \delta_{ij} = E_i + \sum_{\alpha} \frac{1}{2} \omega_{\alpha} Q^2_{\alpha} \quad ; \alpha=1,3,5,7,8,9,12.
\label{eq:a}
\end{alignat}
Higher order terms then enter the appropriate higher order diabatic potential matrices.

Going back to equation \ref{eq:1}, the order of the expansion gives the name for the model: linear vibronic coupling model for expansions truncated at $\bm{W}^{(1)}$, quadratic vibronic coupling model for truncations at $\bm{W}^{(2)}$, quartic vibronic coupling model for truncations at $\bm{W}^{(4)}$, etc. These terms are non-diagonal matrices and are responsible for the coupling, mediated by vibrations, between pairs of electronic states, hence the name of vibronic coupling model. 
In the 1,1-DFE model, we included expansion terms up to sixth order. The terms up to second order are
\begin{align}
	&W^{(1)}_{ii}(Q) = \sum_{\alpha} \kappa^{(i)}_{\alpha} Q_{\alpha} \nonumber \\
	&W^{(1)}_{ij}(Q) = \sum_{\alpha} \lambda^{(i,j)}_{\alpha} Q_{\alpha} \, , \;\, i \neq j \nonumber \\
	&W^{(2)}_{ii}(Q) = \sum_{\alpha, \beta} \frac{1}{2} \gamma^{(i)}_{\alpha,\beta} Q_{\alpha} Q_{\beta} \nonumber \\
	&W^{(2)}_{ij}(Q) = \sum_{\alpha, \beta} \frac{1}{2} \mu^{(i,j)}_{\alpha,\beta} Q_{\alpha} Q_{\beta} \, , \;\, i \neq j
\label{eq:3}
\end{align}
showing how the dependence with the degrees of freedom is linear for the first term, quadratic for the second, etc. Higher-order terms were also
required for the diabatic potentials of some modes, in particular for the on-diagonal (intra-state) mode coupling, and are written
\begin{align}
	&W^{(m)}_{ij}(Q_\alpha) = \frac{1}{m!} C^{m,(i,j)}_{\alpha} Q_{\alpha}^m \nonumber \\
	&W^{(m+n)}_{ij}(Q) = \frac{2}{(m+n)!} C^{m,n,(i,j)}_{\alpha,\beta} Q^m_{\alpha} Q^n_{\beta}.
\label{eq:3a}
\end{align}
Since 1,1-DFE belongs to the C2v symmetry point group we can apply symmetry to
neglect the terms that would make the molecular Hamiltonian antisymmetric.
Since the Hamiltonian must be invariant under the symmetry operations of the C$_{2v}$ symmetry point group, to which 1,1-DFE belongs, some of the Hamiltonian terms are zero by definition. The terms up to second order which are nonzero obey the following product rules for the irreps of the vibrations, $\Gamma_\alpha, \Gamma_\beta$, and the irreps of the electronic wavefunctions, $\Gamma_i, \Gamma_j$
\begin{alignat}{4}
	&\kappa^{(i)}_{\alpha}&& \neq 0 : \;\, \Gamma_{\alpha} \supset \Gamma_A \nonumber \\
	&\lambda^{(i,j)}_{\alpha}&& \neq 0 : \;\, \Gamma_{\alpha} \otimes \Gamma_i \otimes \Gamma_j \supset \Gamma_A \nonumber \\
	&\gamma^{(i)}_{\alpha,\beta}&& \neq 0 : \;\, \Gamma_{\alpha} \otimes \Gamma_{\beta} \supset \Gamma_A \nonumber \\
	&\mu^{(i,j)}_{\alpha,\beta}&& \neq 0 : \;\, \Gamma_{\alpha} \otimes \Gamma_{\beta} \otimes \Gamma_i \otimes \Gamma_j \supset \Gamma_A.
\label{eq:4}
\end{alignat}
This means that the excited state gradients, $\kappa$, are only non-zero for totally symmetric modes, while the linear vibronic coupling between states, $\lambda$, is due to modes with the correct symmetry that when multiplied by the symmetry of the pair of states leads to a totally symmetric irreducible representation. The diagonal $\gamma$ terms will be non-zero if the direct product of both modes involved is totally symmetric, whereas for the off-diagonal quadratic coupling terms, $\mu$, the product must also include the symmetry of the electronic states. Symmetry rules for the higher order terms follow as straighforward extensions of these.

Some modes with strong anharmonicity require diabatic potential functions in place of the Taylor expansions described above in order to go to higher orders without drastically increasing the number of parameters to fit. The off-diagonal coupling, however, always follows the scheme above. These anharmonic functions were chosen purely on how well they fitted the ab initio data. Modes q$_2$ and q$_{11}$ are described by Morse potentials:
\begin{equation}
W_{Mor,ii}(Q_\alpha) =  D^{(i)}_{0} \left\{ 1 - \exp \left[-a_\alpha^{(i)} \left( Q_{\alpha} - Q^{(i)}_{0,\alpha} \right)\right]\right\}^2 + \epsilon^{(i)}_{\alpha}
\end{equation}
and modes q$_6$ and q$_{10}$ used a Lennard-Jones form
\begin{equation}
W_{LJ,ii}(Q_\alpha) =  -D^{(i)}_{0}  \left[  \left( \frac{(\alpha_\alpha^{(i)})^{12}}{Q_{\alpha} - Q^{(i)}_{0,\alpha}}\right)^{12} -\left( \frac{2 (\alpha_\alpha^{(i)})^6}{Q_{\alpha} -Q^{(i)}_{0,\alpha}}\right)^{6} \right] + \epsilon^{(i)}_{\alpha}.
\end{equation}
The torsional q$_4$ mode, as explained above, is replaced by an angle, $\theta$. Potentials for this mode are periodic and were fitted as a function of the cosine of the torsion angle:
\begin{equation}
W_{ii}(\theta) =
 \kappa^{(i)}_{4}\mathrm{cos}(\theta) + \frac{1}{2} \gamma^{(i)}_{4} \mathrm{cos}^2(\theta),
\label{eq5}
\end{equation}
where $\kappa^{(i)}_{4}$ and $\gamma^{(i)}_{4}$ are fitting constants. Any diabatic coupling and higher order terms coupling q$_4$ to the other modes used expressions of the form in Eqs. (\ref{eq:3}) and (\ref{eq:3a}), using $\cos(\theta)$ in place of $Q$, and taking into account that this function is symmetric at $\theta=0$ and anti-symmetric at $\theta=\frac{\pi}{2}$. 

\begin{figure}
\centering
 \includegraphics[width=0.7\textwidth]{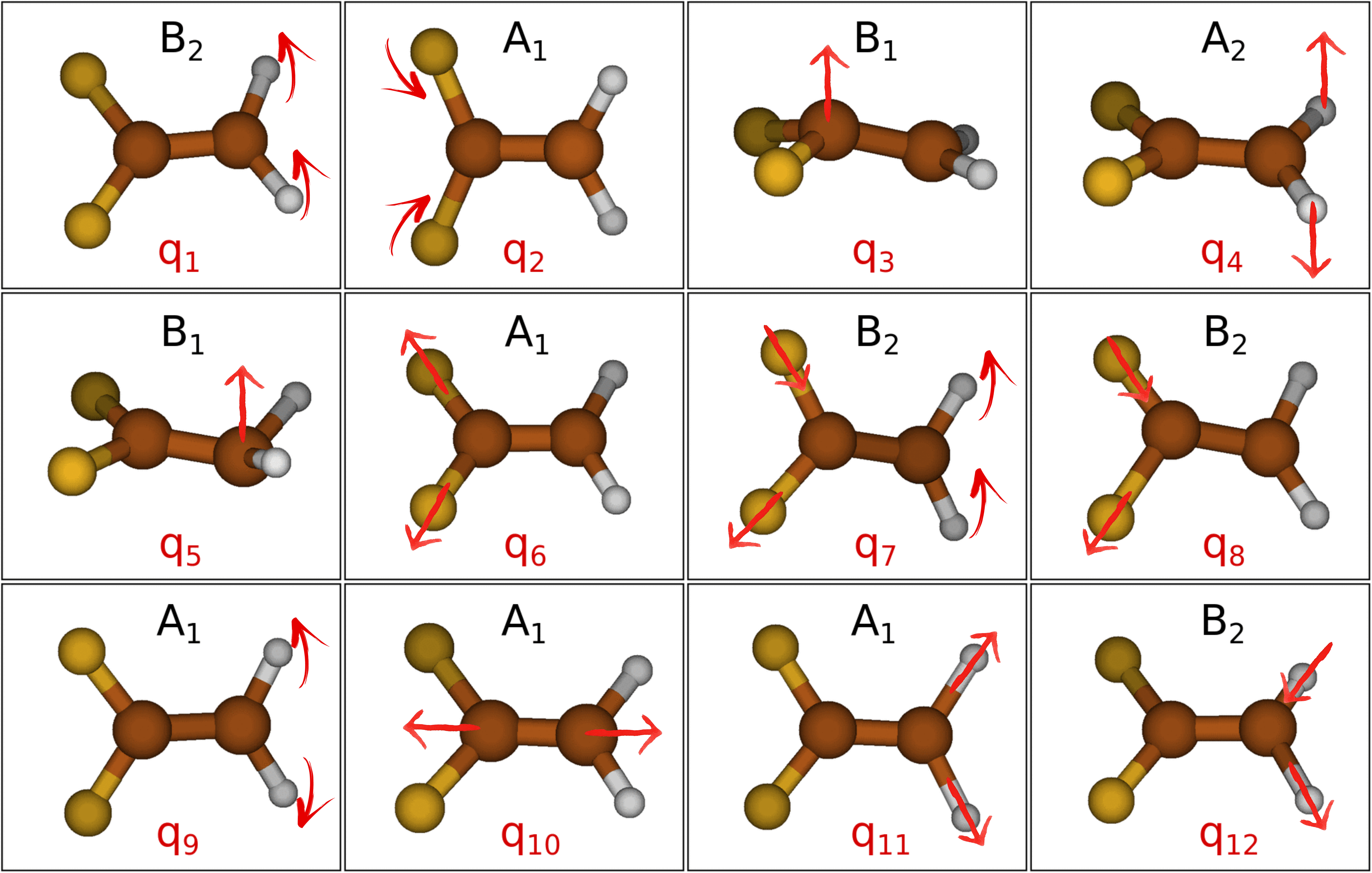}
 \caption{Normal mode coordinate basis with symmetry labels belonging to the C$_{2v}$ point group. The torsional normal mode q$_4$ is replaced by the full torsion $\theta$ in the model Hamiltonian.}
 \label{M1}
\end{figure}

The parameters for the functions were obtained using the VCHam tool in the Quantics suite\cite{Quantics_program} to calculated ab-initio data.
The 1,1-DFE normal modes were obtained at the planar D$_\mathrm{2h}$ optimised ground state geometry at the MP2/aug-cc-pVDZ level of theory using the Gaussian 09 program.\cite{Gaussian2009} The normal modes and the corresponding symmetry labels using the C$_\mathrm{2v}$ point group are depicted in Figure~\ref{M1}. We use C$_\mathrm{2v}$ labels as this point group is retained throughout rotation around the torsional axis. A comparison between the optimised MP2 geometrical parameters and the experimental values is shown in Table~\ref{table1}. A table with frequencies and description of the motions can be found in the Supporting Information in SI-Table 1.

\begin{table}
\centering
\caption{Bond lengths (in {\textup{\AA}}ngstrom) and angles (in degrees) for the geometries obtained bu optimisation of the ground state of 1,1-DFE employing MP2/aug-cc-pVDZ in comparison with experimental data from Ref.~\citenum{exp1975}.}
\begin{tabular}{|l|cc|}
\hline
& Exp\cite{exp1975} & MP2 \\
\hline
r(CC)& 1.315 & 1.335 \\ 
r(CF)& 1.323 & 1.337 \\
r(CH)& 1.078 & 1.087 \\
$\sphericalangle$(FCF) & 109.1& 109.6\\
$\sphericalangle$(HCH) & 121.8& 121.5 \\
$\sphericalangle$(CCF)& 125.5& 125.2\\
$\sphericalangle$(HCC) & 119.1 &119.3\\
\hline
\end{tabular}
\label{table1}
\end{table}

The electronic state energies of the lowest seven singlet states at torsion angles of
0$^\circ$ and 90$^\circ$ calculated at the SA(11)-MS-CAS(2,6)PT2/aug-cc-pvDZ  level of theory are listed in Table \ref{table2}. The active space includes the Rydberg orbitals 3s, 3p$_x$, 3p$_y$, and 3p$_z$ and the $\pi$ and $\pi^*$ orbitals. The state-average procedure is performed over eleven roots to account for every single and double excitation related to the chosen active space (as explained in Ref.\citenum{gomez2019}), even if here we were only interested in the dynamics in the lowest seven states. A picture of the active space and further details about the electronic structure can be found in Ref.~\citenum{gomez2019}.

\begin{table}
\centering
  \caption{Vertical excitations ($\Delta$E, in eV) of 1,1-DFE at the SA(11)-MS-CAS(2,6)PT2/aug-cc-pvDZ level of theory. Three electronic states are labelled following the Mulliken notation for ethylene,\cite{mulliken1933} N for the closed shell $\pi^2$, V for the $\pi\pi^*$ and Z for the $\pi^{*2}$ state. The energies are calculated at the optimised geometry (FC, for Frank Condon) and at 90$^\circ$ of torsion. The experimental (exp) values are taken from Ref.~\citenum{limao2006}.}
  \label{table2}
\begin{tabular}{|lcccc|}
\hline
State & $\Delta$E at FC & exp\cite{limao2006} & symmetry & $\Delta$E at 90$^\circ$\\ 
\hline
N               & 0.00 & & A$_1$  & 3.30  \\
V               & 7.45 &7.49 & A$_1$ &  3.47 \\
$\pi$-3s              & 7.07 & 7.06 & B$_2$ & 7.91 \\
$\pi$-$\mathrm{3p_x}$ & 8.10 & & A$_2$ & 9.03 \\
$\pi$-$\mathrm{3p_y}$ & 8.20 & & B$_1$ & 9.26 \\
$\pi$-$\mathrm{3p_z}$ & 8.93 & 8.97 & A$_1$ & 10.03 \\
Z               & 13.77& & A$_1$ & 9.27 \\
\hline
\end{tabular}
\end{table}

\begin{figure} 
\centering
    \includegraphics[width=\textwidth]{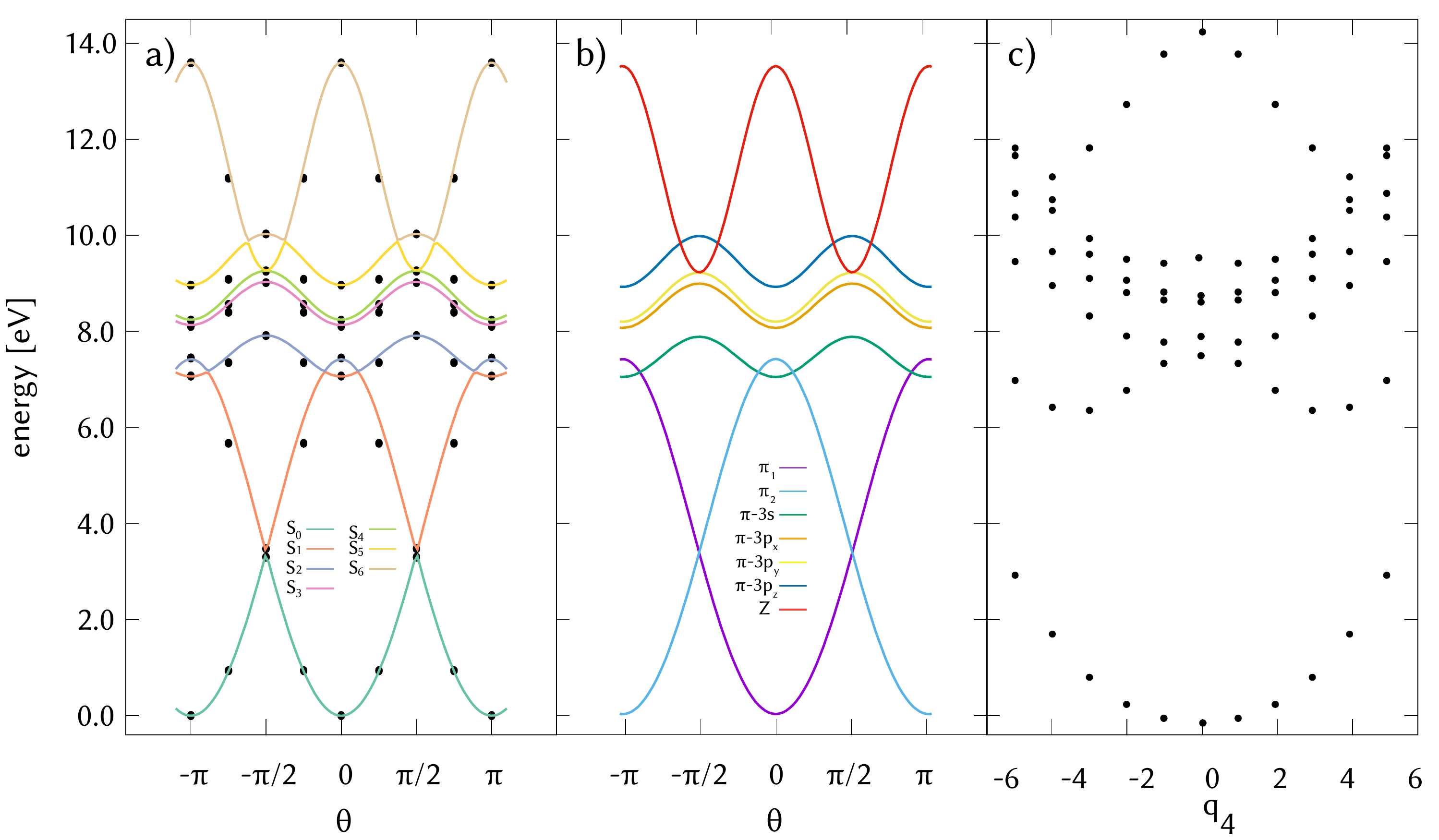}
\caption{Potential energy curves as a function of the torsional angle $\theta$ of 1,1-DFE with all other coordinates at the equilibrium geometry. Fit made using the VCHam tool in the Quantics\cite{Quantics_program} suite to ab-initio points calculated with SA(11)-MS-CAS(2,6)PT2/aug-cc-pvDZ. a) Adiabatic curves and ab-initio points. b) Diabatic curves. The orbital-diabatic potentials $\pi_1$ and $\pi_2$ (purple and blue respectively) cross at $\pi$/2 radians and have two equivalent minima at 0 and $\pi$ radians, each one from a different diabat. c) Ab-initio calculated energy points as a function of the torsional normal mode $q_4$.}
        \label{M2}
\end{figure}

The full dimensional potential energy surface consists of 12 nuclear degrees of freedom and 7 electronic states: N, V, Z, $\pi$-3s, $\pi$-3p$_x$,  $\pi$-3p$_y$, and  $\pi$-3p$_z$. N stands for neutral, V for valence and Z for zwitterionic using the Mulliken notation for ethylene\cite{mulliken1933}. These labels represent the character of the wavefunction at the Franck-Condon point and are thus used to label the diabatic states of our model.  Since we need two minima equivalent by torsion to construct the initial wavefunction as a torsional superposition, we transformed our potentials using an orbital diabatic basis. In this basis, the N and V potentials cross and interchange character (see Figure~\ref{M2}b). In this new basis, we rename the states as $\pi_1$ and $\pi_2$, since they are now completely equivalent. The rest of the states mentioned above keep their original character and thus their labels do not change. 
The 7 diabatic potential energy curves along the torsion were fitted to ab-initio calculations using the SA(11)-MS-CAS(2,6)PT2/aug-cc-pVDZ level of theory performed at different geometries. Adiabatic curves and single ab-initio points are plotted in Figure~\ref{M2}a. The corresponding diabats are plotted in Figure~\ref{M2}b. 
The potentials along the original normal mode q$_4$ can be seen in Figure~\ref{M2}c\, showing the need of transforming the torsional coordinate in order to describe the crossing to the ground state. The two lowest  states cross at 90 degree of torsion.

To provide data for the fitting, calculations at the FC geometry and at 90 degree of torsion were performed along every normal mode providing 3234 data points, corresponding to 462 single point electronic structure calculations. The model has 553 parameters to fit which are not independent and so the fitting was done in a series of steps to allow control of the procedure. The diabatic potentials for the torsional degree of freedom were fitted first, using the planar geometry at 0 degree as the initial point for the Taylor expansion. Next, for every normal mode a 1D fit to the ab-initio data was performed, starting the expansion at 90 degrees of torsion. Explicit couplings to the torsional degree of freedom were then included to force the two minima at planar geometries (0 and 180 degree) to be equivalent. For every 1D fit, the configuration interaction vector of the electronic structure calculation was monitored, to ensure that the diabats correctly follow their electronic character. For the states $\pi_1$ and $\pi_2$, however, the same diabatic functions must be used to impose symmetry, which caused some deviations between the ab-initio data and the fitted curves.


\begin{figure} 
\centering
    \includegraphics[width=0.71\textwidth]{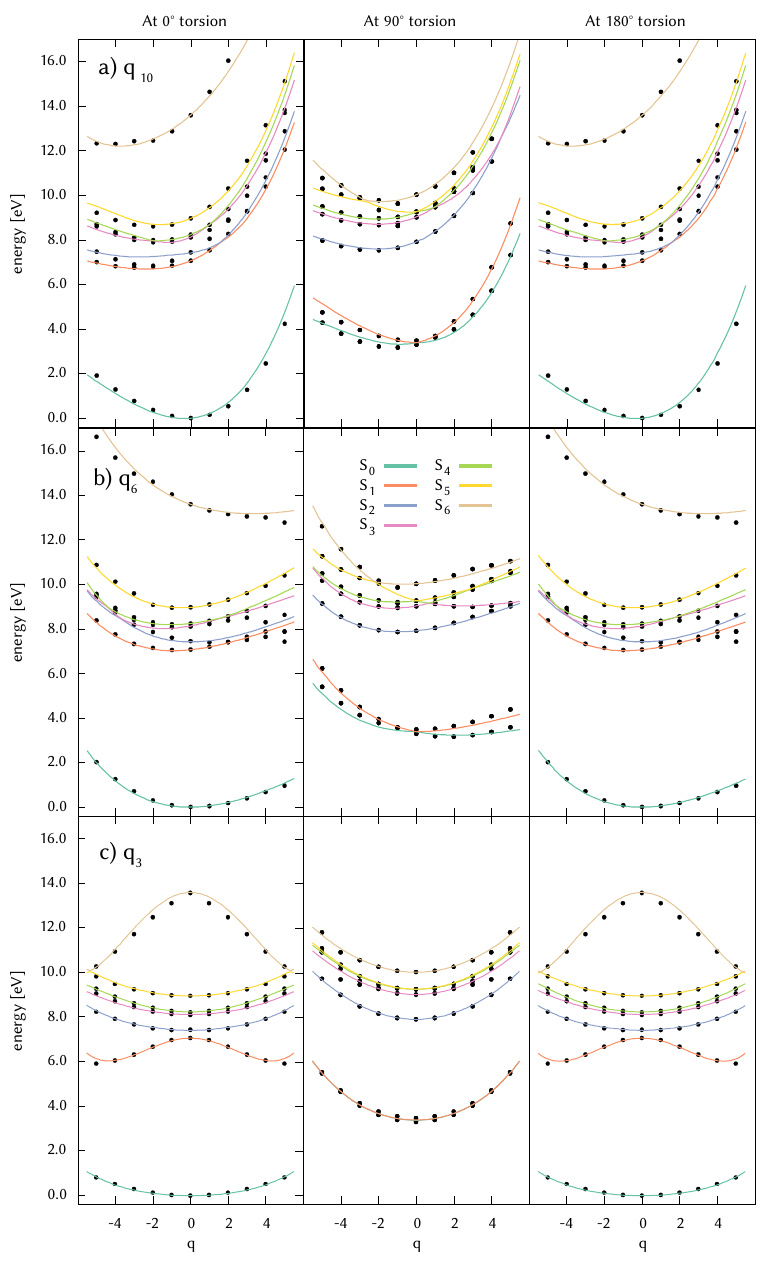}
\caption{Adiabatic potential energy curves along key normal mode coordinates of 1,1-DFE calculated with SA(11)-MS-CAS(2,6)PT2/aug-cc-pvDZ. The right and left columns correspond to torsional angles of $\pi$ (180 degree) and 0 degrees, respectively. The middle column corresponds to a torsional angle of $\pi$/2 (90 degree), where the fit was performed. All other coordinates are at equilibrium values. The labels q$_i$ correspond to the normal mode coordinate along which the ab-initio energies were calculated and fitted. The corresponding normal modes can be seen in Figure~\ref{M1}.}
        \label{M4}
\end{figure}

 After the initial guesses for every diabatic 1D potential were found, a global fit was performed including couplings up to sixth order for specific coordinates (especially couplings that involved the torsional motion). The total standard deviation with respect to the ab-initio points was $0.257$\,eV, with the largest errors being associated with high energy points that are not well fitted by the model potentials and do not play a role in the dynamics. The 1D adiabatic potential energy curves and the corresponding ab-initio points for three key modes ($q_3, q_6$ and $q_{10}$) are plotted in Figure~\ref{M4}. Cuts along the other degrees of freedom can be found in the Supporting Information. The middle column shows the potential curves at 90 degree of torsion, where the fit was performed. Note that to its right and its left, at both 0 and 180 degrees of torsion, the potentials are equivalent, conserving the symmetry. However, these two minima correspond to different diabatic potentials ($\pi_1$ and $\pi_2$) which switch energetic order at 90 degree. To preserve the symmetry, these diabatic potentials thus need to be fitted with the same analytical functions and parameters for each normal mode.

All parameters obtained for the model Hamiltonian are listed in the Supporting Information, along with cuts like that in Fig.~\ref{M4} for all modes showing the quality of the fit.

\subsection{Initial conditions}

\begin{figure} 
\centering
    \includegraphics[width=0.5\textwidth]{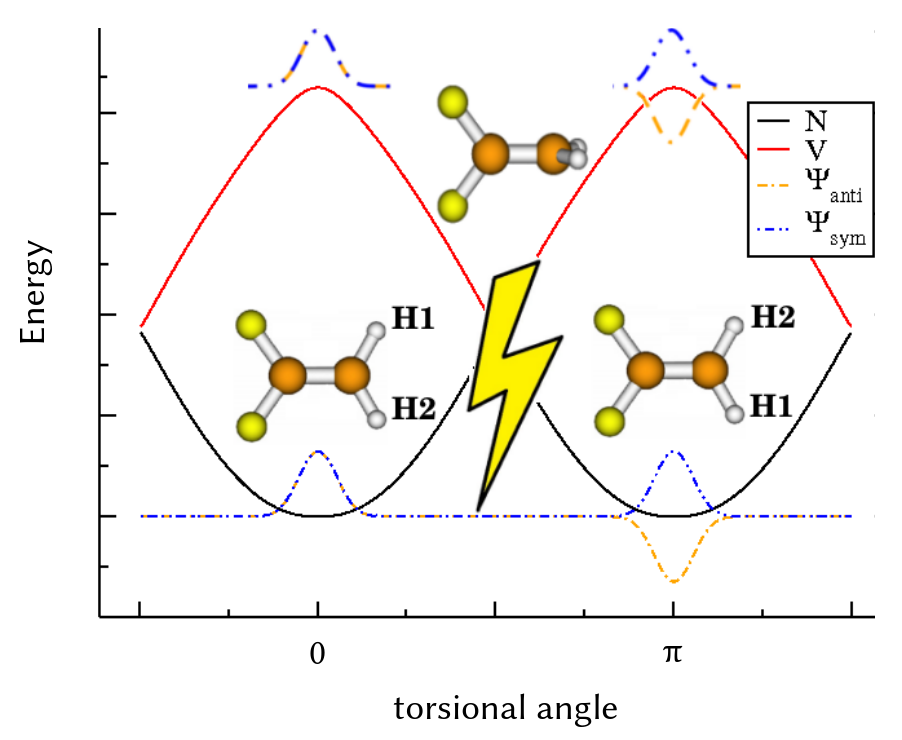}
\caption{Scheme depicting the initial conditions used in this work. Either the symmetric (blue), the antisymmetric (orange) or the localised (just the eigenstate of the left side minima with positive amplitude) were vertically excited to initiate the dynamics.}
        \label{initconds}
\end{figure}

Calculations were performed on different size systems (2D, 5D and 12D). The initial wavefunction in each calculation was constructed as the Hartree product of the harmonic oscillator ground state for every normal mode and the torsional ground state wavefunction in the $\pi_1$ minimum. This was followed by propagation in imaginary time (energy relaxation)\cite{Auer2001} to get the ground-state vibrational eigenfunction. The barrier between the minima is high enough that on the timescale of the relaxation (100 fs) there is no transfer of population to the $\pi_2$ minimum. Three different initial conditions for the torsional wavefunction were then constructed from this localised wavefunction obtained by relaxation; taken directly as the eigenstate of the $\pi_1$, or as positive and negative superpositions to form approximate eigenstates of the $\pi_1$ and $\pi_2$ diabatic potential energy curves. A scheme can be seen in Figure~\ref{initconds}.

This multidimensional initial wavefunctions created in this way for one or two torsional minima corresponding to states $\pi_1$ and $\pi_2$ was consequently stored, and read in to be further electronically excited to the $\pi_2$ and $\pi_1$ states, respectively. This way, either a localised torsional wavefunction or a superposition of them was created in the S$_2$ adiabatic state (blue state in  Figure~\ref{M2}a). In the case of the localised initial condition, the wavefunction carried the whole population, while for the superposition each half was given an amplitude of $1/\sqrt{2}$.

\section{Results}  
\subsection{Dynamics started on the $\pi\pi^*$ valence state}

Quantum dynamics simulations of 1,1,-DFE were performed for symmetric and antisymmetric superpositions as well as for a system localised in one single minima. 

For each case, a propagation using the MCTDH method was performed for $150$\,fs,  using the multiset formalism, i.e., different single particle functions (SPFs) are optimised for each electronic state. The mean fields were calculated at each integration step and integrated with the Adams-Bashforth-Moulton\cite{Burden1993} predictor-corrector method. 
The torsional coordinate was described using a Fast Fourier Transform (FFT) collocation grid, while the other degrees of freedom used harmonic oscillator discrete variable representations (DVRs) as primitive basis functions. 

Since convergence of the MCTDH method depends on how the modes are combined within a SPF, on the number of them and on the primitive grid size, we first started converging calculations on a reduced system, to steadily increase the number of degrees of freedom. We performed calculations with two (torsion $\theta$ and CC stretching q$_{10}$ ), five (torsion $\theta$, CC stretching q$_{10}$, CF$_2$ pyramidalisation q$_{3}$, CF symmetric stretch q$_{6}$ and a combination of asymetric CF$_2$ stretch and CH$_2$ bend, q$_{7}$) and twelve degrees of freedom, i.e., full dimensionality (the motions are depicted in Figure~\ref{M1}). Calculations for the three different initial conditions in all sized systems (2, 5, or 12 degrees of freedom) used the same primitive grids, but were converged individually with respect to the SPF basis. The number of primitive basis functions, along with mode combinations and the number of SPFs per state for the largest calculations performed is found in Table~\ref{basis-qua}. 

\begin{table}
\begin{tabular}{|c|cccc|}
\hline
Type of WF & particle & $N_i$ & $n_{i}$ & Config.\\
\hline
 {\bf 2mode} & & & & \\ 
  local & el & 7 &  & 7\\
    & $\theta$,q$_{10}$  & 100,55 & (1,1,1,1,1,1,1) & \\
    & & & & \\
 sym / anti & el & 7 &  & 7\\
      & $\theta$,q$_{10}$  & 100,55 & (1,1,1,1,1,1,1)& \\
\hline
 {\bf 5mode} & & & & \\
  local & el & 7 &  & 35278\\ 
   & $\theta$,q$_{7}$  & 100,27 & (10,29,26,8,3,3,4)&  \\
   & q$_{10}$ & 55 & (16,20,20,7, 3,3,4)& \\
   & q$_3$,q$_6$ & 31,27 & (10,30,20,10,3,3,4) & \\
   & & & & \\
 sym / anti & el & 7 &  & 54269\\
   & $\theta$,q$_{7}$  & 100,27 & (23,32,32,10,4,3,4) & \\
   & q$_{10}$ & 55 & (19,23,28,8,3,3,4)& \\
   & q$_3$,q$_6$ & 31,27 & (10,30,30,10,4,3,4) & \\
\hline
{\bf 12mode} & & & & \\ 
  local & el & 7 &  & 229112\\
  &  $\theta$,q$_{10}$  & 100,55 & (17,27,37,11,4,3,3)& \\
  & q$_1$,q$_2$,q$_5$,q$_8$ & 11 & (5,9,10,6,3,3,3)& \\
  &  q$_3$,q$_6$,q$_7$ & 31,27,27 & (17,29,30,9,4,3,4)& \\
 & q$_9$, q$_{11}$,q$_{12}$ & 11 & (10,11,12,6,4,3,3)& \\
   & & & & \\
 sym / anti &  el & 7 &  & 619117 \\
   & $\theta$,q$_{10}$  & 100,55 & (30,40,40,20,4,3,4)& \\
  & q$_1$,q$_2$,q$_5$,q$_8$ & 11 & (11,10,12,9,3,3,3)& \\
 &  q$_3$,q$_6$,q$_7$ & 31,27,27 & (30,40,40,20,4,3,4)& \\
 & q$_9$, q$_{11}$,q$_{12}$ & 11 & (13,13,13,9,4,3,3)& \\
\hline
\end{tabular}
\caption{Basis used for the largest MCTDH calculations starting with localised and delocalised (symmetric and antisymmetric) wavefunctions on the $\pi\pi^*$ state. The primitive basis was the harmonic oscillator DVR\cite{light1985} for every mode except the torsion, which used a periodic FFT grid.\cite{kosloff1983}  $N_i$ is the number of primitive basis functions for each degree of freedom in the particles, while $n_{i}$ is the number of SPFs used for each of the seven electronic states. Config is the total number of configurations in the MCTDH wavefunction.} 
\label{basis-qua}
\end{table}

\begin{figure}
\centering
    \includegraphics[width=0.8\textwidth]{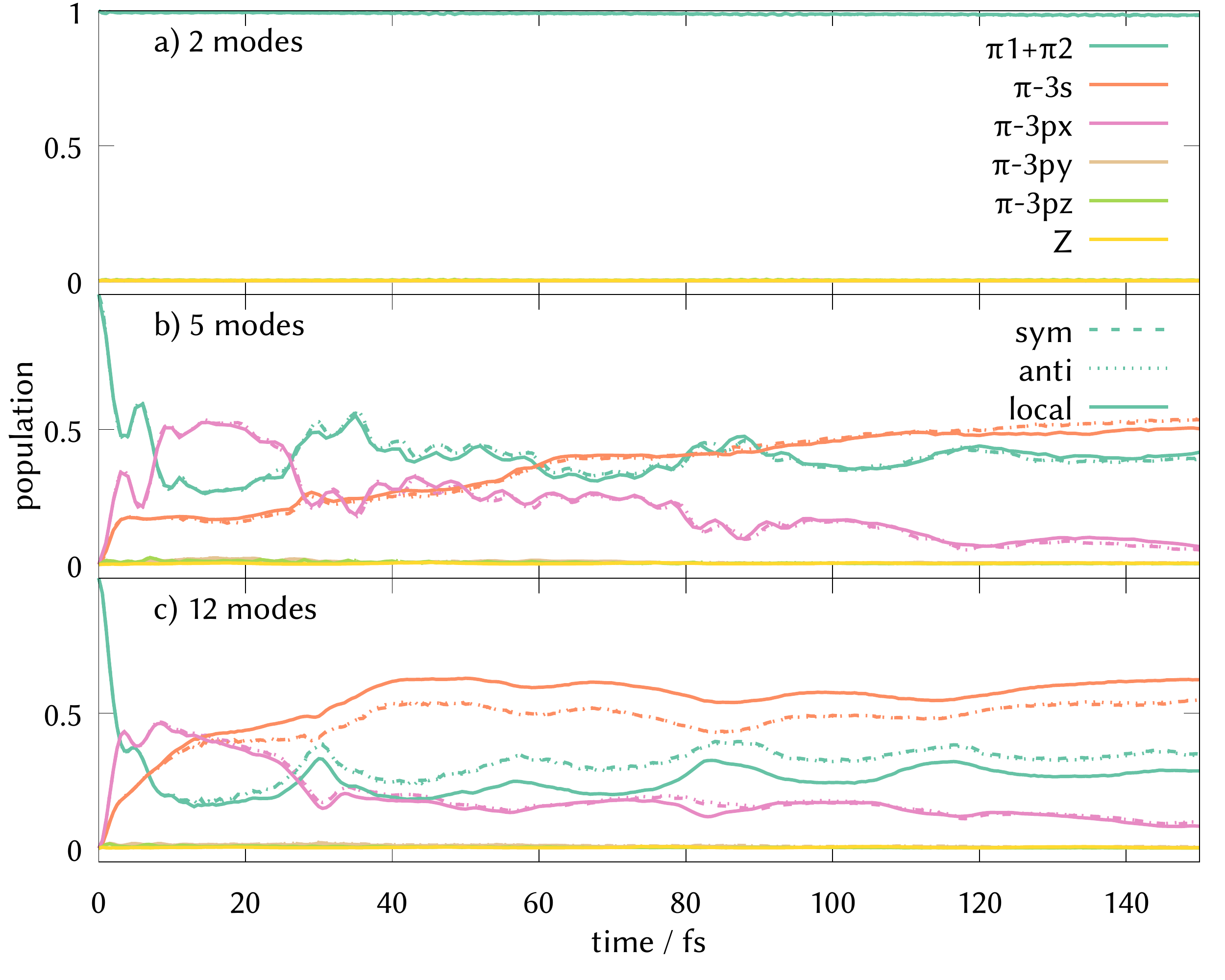}
\caption{Diabatic time-dependent electronic state populations for calculations including (a) two nuclear degrees of freedom: torsion and CC stretch, (b) five nuclear degrees of freedom: torsion, CC stretch and scissoring, and pyramidalisation of CH$_2$ and CF$_2$ fragments, (c) twelve degrees of freedom, i.e., full dimensionality. In solid lines the population after exciting a localised initial wavefunction (local) is depicted. In dashed and in dotted lines the populations after exciting the symmetric superposition (sym) and the antisymmetric (anti), respectively, are shown. }
\label{M8}
\end{figure}

The dynamics including only two degrees of freedom, q$_{10}$ and $\theta$, shows no interchange of population between the $\pi_1$-$\pi_2$ states and the $\pi$-3p$_x$ state during $150$\,fs  (Figure~\ref{M8}a), indicating that the CC double bond vibration is not enough to drive the system away from the Frank-Condon region or excite the torsional motion to reach the conical intersections. This agrees with previous work on ethene which showed that for population transfer to occur the pyramidalisation motion also needs to be included \cite{BENNUN2000-2}
The populations of the $\pi_1$ and $\pi_2$ states have been summed up in the superposition calculations to allow comparison with the localised initial condition that started the dynamics with 100\% of its population in $\pi_2$. The differences of using or not using a superposition as initial condition are unnoticeable. As we can see in Figure~\ref{M8}b, at least five modes are needed to drive the dynamics, which are qualitatively similar to the full dimensionality result shown in Figure~\ref{M8}c. The population, initially excited to the V state, is very quickly transferred to the $\pi$-3s and $\pi$-3p$_x$ Rydberg states. This transfer is initially mediated by q$_7$, a combination of asymmetric CF$_2$ stretch and CH$_2$ bend, and the torsional degree of freedom q$_{4}$ with the same symmetry as the electronic states and it is driven by the gradient created by the CF symmetric (q$_6$) and the CC double bond (q$_{10}$) stretches. In the 5D and 12D cases, the $\pi$-3p$_x$ state traps the population during the first $30$\,fs, after which it slowly relaxes to the $\pi_1$-$\pi_2$ states and the $\pi$-3s. In the 12D case (Figure~\ref{M8}c), the population transferred to the $\pi$-3s state is larger than in the reduced dimensionality system, indicating that other degrees of freedom play a role in the excited state dynamics. When comparing the localised and delocalised initial condition 12D runs, we can observe small differences in the populations, probably due to the difficulty of converging the SPF basis in the larger systems. 

\begin{figure}
\centering
    \includegraphics[width=0.65\textwidth]{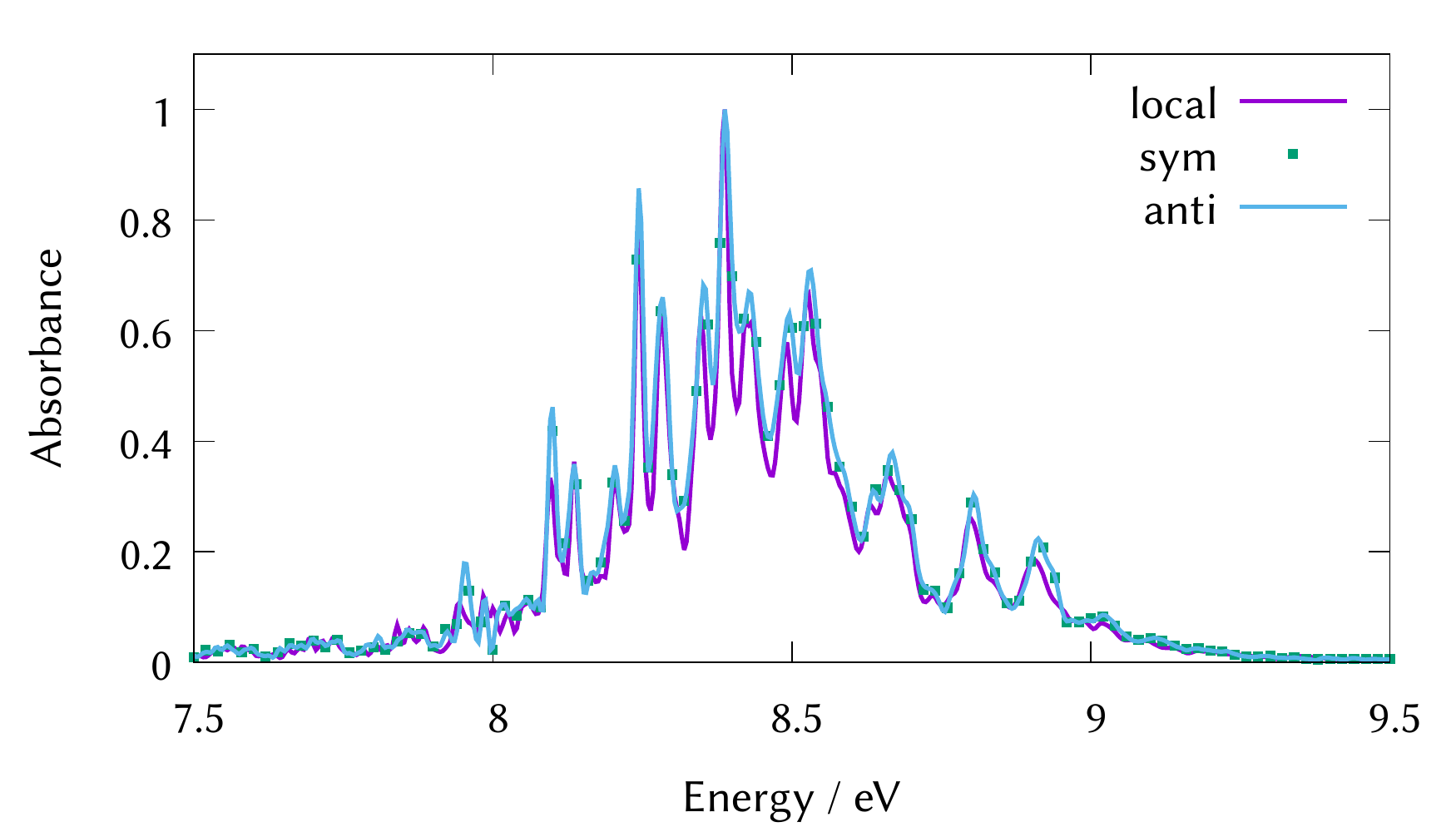}
\caption{Absorption spectra calculated as the Fourier transform of the autocorrelation function from full dimensionality  dynamics with a damping time (phenomenological broadening) of $\tau=$ 100 fs comparing an initial condition localised in one minimum (local) a symmetric superposition (sym) and an antisymmetric superposition (anti) started on the V state. }
        \label{M9}
\end{figure}


During the dynamics, the autocorrelation function (overlap of wavefunctions at time\,=\,t and time\,=\,0) was stored every $0.5$\,fs and later Fourier transformed to obtain the absorption spectrum. In Figure~\ref{M9}, the spectrum for the localised initial condition (local) and superpositions with positive (sym) and negative (anti) relative phase are shown for the full system with twelve degrees of freedom. The minimal differences that can be observed are again probably due to lack of basis set convergence. 
The main spacing is due to the CC stretching mode (q$_{10}$) with a harmonic frequency of $0.2192$\,eV and the shoulder peaks next to the main progression are due to the CF$_2$ symmetric stretch (q$_6$, harmonic frequency 0.1132 eV). By comparing to the spectra from reduced dimensionality calculations, the overall width and shape comes from the torsional degree of freedom and the CF symmetric stretches which couple the electronic states, driving the initial wavepacket away from the Frank-Condon region. 

\begin{figure}
\centering
    \includegraphics[width=\textwidth]{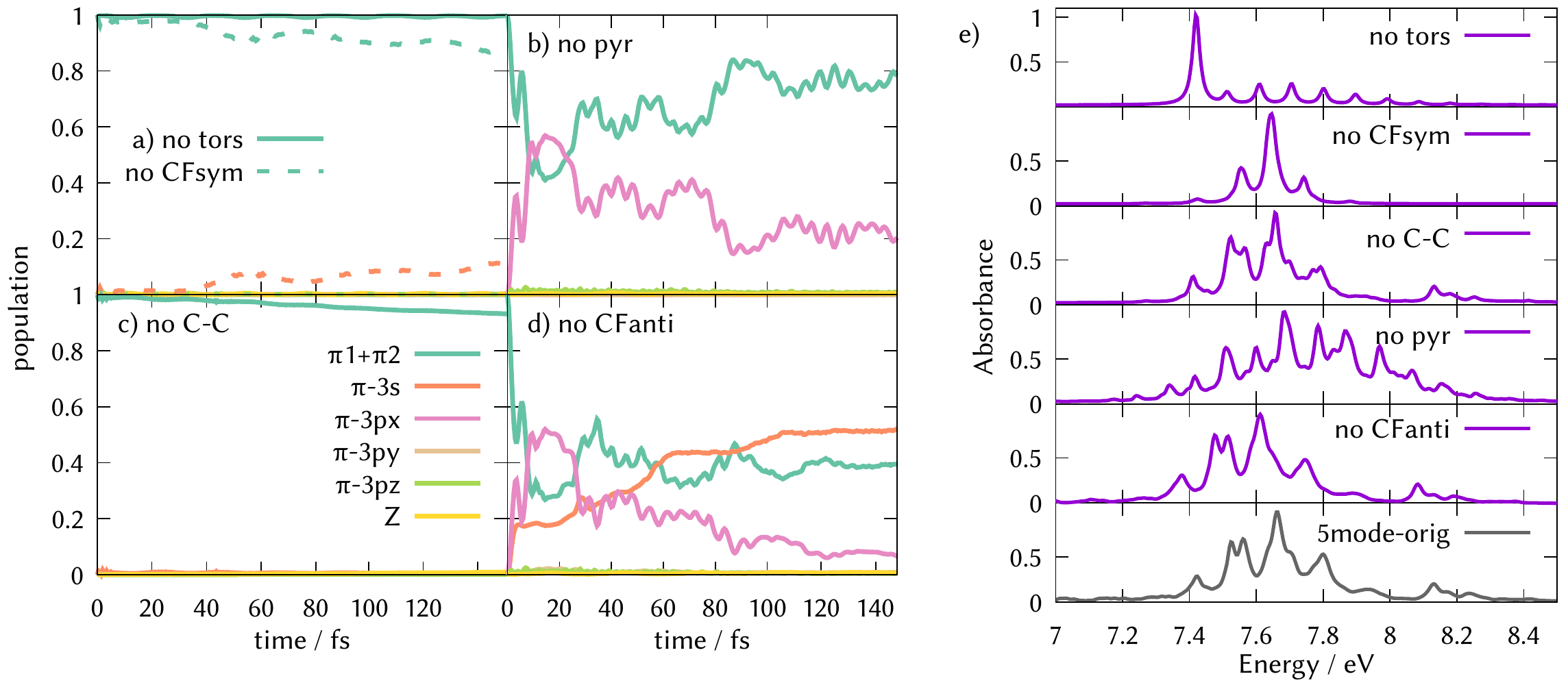}
\caption{Simulations corresponding to the five mode dynamics in figure \ref{M8} with a localised initial condition, after removing one degree of freedom. a) Diabatic time-dependent populations after removing the torsional $\theta$ mode (solid line) and after removing the CF symmetric stretch q$_6$ (dashed), b) without including the CF$_2$ pyramidalisation q$_3$, c) ignoring the CC stretch mode q$_{10}$ and d) ignoring q$_7$, a combination of asymmetric CF$_2$ stretch and CH$_2$ bend. e) Absorption spectra computed as the Fourier transform of the autocorrelation function for each of these cases.}
\label{M10}
\end{figure}

In contrast to a linear vibronic coupling model, our high-order Hamiltonian makes the assignment of which modes couple which states and drive the dynamics impossible after a few femtoseconds. To aid in this task, we performed excited state dynamics based on the 5D propagation shown in Figure~\ref{M8}b, but using only the localised initial condition and removing one degree of freedom each time. The populations and the absorption spectra are depicted in Figure~\ref{M10}. These simulations show clearly that the torsion and the CF symmetric and CC stretches are crucial to obtain population transfer to any state on the ultrafast timescale, since removing any of these DoFs inhibits the deactivation of the bright state. The CF$_2$ pyramidalization is needed to allow a transfer to the  $\pi$-3s state. Although q$_7$, a combination of asymmetric CF$_2$ stretch and CH$_2$ bend, was initially thought to couple the valence and the $\pi$-3s state for symmetry reasons, its presence does not affect the outcome of the dynamics. The spectra shown in Figure~\ref{M10}e supports the assumption that the torsion dominates the spectrum, since without it we obtain a vibrational progression where the maximum is the 0-0 band. The CF symmetric stretch is responsible for the shoulder peaks and the pyramidalization narrows down the general shape of the spectra when allowing the transfer to the $\pi$-3s state. 

In our previous work on 1,1-DFE using on-the-fly dynamics\cite{gomez2019}, the $\pi$-3s state acted as a doorway state, keeping population from the V state trapped for a while before transferring it back. This process was essential for the deactivation of the molecule to its electronic ground state. Using our parametrised potentials and quantum dynamics, we do not see any trapping of the population on this Rydberg state. Instead, its population keeps growing and after 150 fs it is the most populated state. To distinguish between the $\pi_1$-$\pi_2$ state populations we calculated the adiabatic populations for the 5D localised system integrating the transformed wavefunction over the primitive grids. The electronic ground state population was found to never exceed the value of 0.6\%. Therefore, the outcome of the dynamics is substantially different in the two cases and deserves a closer look why this could be.


A quantum dynamics method (like MCTDH) is fully quantum and thus expected to  provide a better description of the nuclear motion than that obtained by propagating independent classical trajectories, as done in surface hopping. An important difference in the two calculations is that in surface hopping, the trajectories were selected from a Wigner distribution, which is a quasiprobability function that provides an approximation for the lowest eigenstate of the quantum harmonic oscillator; instead, the MCTDH simulations use the actual lowest energy ``localised'' eigenfunction of the Hamiltonian for the initial wavefunction. However, to use the MCTDH method, one must pre-compute the potential energy surfaces in advance and fit them to an analytical potential form on a base of nuclear coordinates. This procedure potentially restricts the system as the choice of coordinates and the fitted function may not allow the system to evolve in the same way as with on-the-fly dynamical simulations. As an example, one could think of the different MCTDH dynamics that would result if we had used the torsional normal mode in the model potential and not substituted it by the full dihedral angle, as depicted in Figure~\ref{M2}. While the quality of the potential used in the MCTDH simulations is supported by the calculated spectrum, the latter only samples the potential energy surfaces around the Frank-Condon point and does not say anything about the surfaces in general. Furthermore, due to its construction the vibronic coupling model does not include any coupling between modes that requires high-order polynomials.

Further work is thus needed to see which of the two sets of dynamical simulations is correct. A possible protocol would be to run on-the-fly quantum dynamics   with the DD-vMCG method \cite{chr21:124127,ric17:606}, but this would require calculating frequencies, gradients, couplings and energies at the CASPT2 level for seven states and each pair of states, which would be computationally very expensive.

\subsection{Localised dynamics started on the Rydberg states}

\begin{figure}
\centering
    \includegraphics[width=0.7\textwidth]{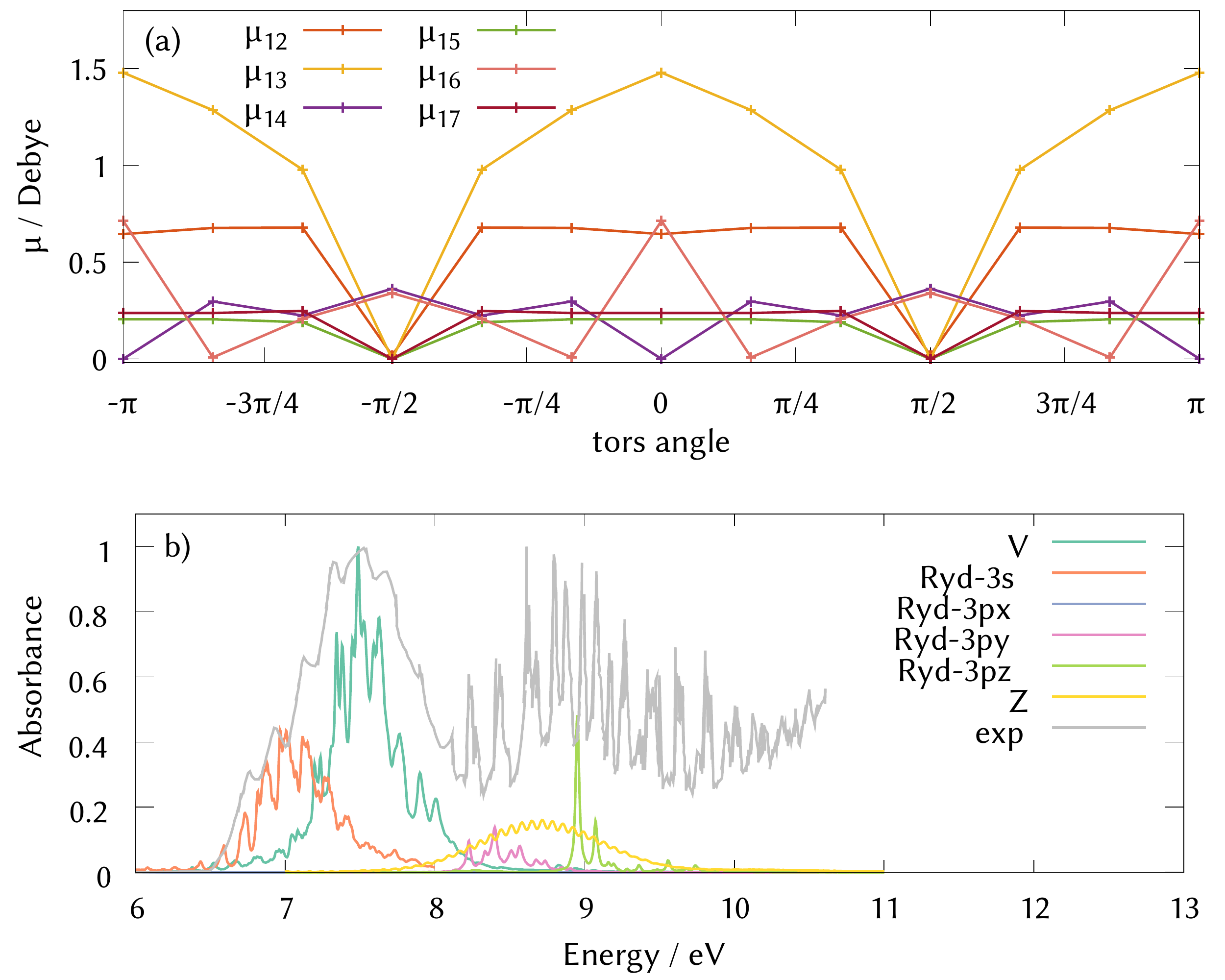}
\caption{a) Elements of the adiabatic state transition dipole moment matrix along the torsional degree of freedom calculated with MS-CASPT2. The sub-indices reflect the pair of electronic states involved in the transition. b) Absorption spectra calculated as the Fourier transform of the autocorrelation function from dynamics including five degrees of freedom (except for the V case, where the full dimensionality result was taken), as specified in Table~\ref{basis-qua2} with a damping of $\tau=$ 50\,fs. Separate calculations were made used a localised initial wavefunction started on the valence, zwitterionic and Rydberg electronic states. The amplitude has been weighted by the transition dipole moments. The experimental spectrum shown in grey was taken from Ref.~\citenum{limao2006}. }
\label{M11}
 \end{figure}

In order to extend our excited state dynamics study beyond the the valence $\pi\pi*$ state, we investigate the dynamics of the system including the manifold of electronic states that cross, couple and affect the behaviour after light irradiation. 
In Figure~\ref{M11}a, the transition dipole matrix elements that couple the ground state and excited state electronic wavefunctions are plotted at different torsional angles. The values, calculated at the same level of theory as the potentials, refer to the adiabatic electronic states. However, at the ground state optimised geometry the V and $\pi$-3s states switch character (see for example Figure~\ref{M2}) and we accommodated that change switching the values of the matrix elements $\mu_{12}$ and $\mu_{13}$ at the planar geometry. Since at the Frank-Condon region the transition dipole moment only has negligible values for the coupling with the $\pi$-3p$_x$ state, we started 5D dynamics using the localised initial condition on every other electronic state and calculated the absorption spectra from the autocorrelation function (Figure~\ref{M11}b). Every spectrum has been weighted by the corresponding matrix element of the transition dipole moment at the FC point. The spectrum starting in the V state has been computed from dynamics in full dimensionality (it is the same as Figure~\ref{M9}) and therefore has been shifted in energy by the difference in the zero point energies due to the extra degrees of freedom. In Ref. \citenum{limao2006}, the first band is concluded to be a mixture of the Rydberg $\pi$-3s and the valence $\pi\pi* $ states. The vibrational progression observed at higher energies starts with peaks from the $\pi$-3p$_y$ state, followed by the $\pi$-3p$_z$ and probably higher lying Rydberg states. The Z state modulates the intensity of these peaks with a broad gaussian-like shape.

\begin{table}
\begin{tabular}{|c|cccc|}
\hline
Initial state & particle & $N_i$ & $n_{i}$ & Config.\\
\hline
$\pi$-3s & el & & 7 & 7130\\ 
&  $\theta$,q$_7$  & 100,27 & (5,12,15,3,3,3,3) &\\
& q$_{10}$  & 55 & (10,14,12,4,4,3,3) &\\
&  q$_3$,q$_6$ & 10,27 & (7,20,20,4,4,3,3) &\\
\hline
$\pi$-3p$_x$ & el & & 7 & 46519\\ 
&  $\theta$,q$_7$  & 100,27 & (12,32,32,10,4,3,3) &\\
&   q$_{10}$  & 55 & (19,22,23,9,4,3,4) &\\
&  q$_3$,q$_6$ & 10,27 & (10,30,30,10,4,3,4) & \\
\hline
$\pi$-3p$_y$ & el & & 7 & 45105\\ 
&  $\theta$,q$_7$  & 100,27 & (9,32,31,10,4,3,4) &\\
&  q$_{10}$  & 55 & (16,22,23,10,4,3,4) &\\
&  q$_3$,q$_6$ & 10,27 & (10,30,30,10,4,3,4) & \\
\hline
$\pi$-3p$_z$ & el & & 7 & 26490\\ 
&  $\theta$,q$_7$  & 100,27 & (7,13,28,10,4,4,3) &\\
&  q$_{10}$  & 55 & (11,19,25,10,4,3,4) &\\
&  q$_3$,q$_6$ & 10,27 & (10,23,27,10,4,3,4) & \\
\hline
Z & el & & 7 & 68355\\ 
&  $\theta$,q$_7$  & 100,27 & (24,32,32,10,,4,3,4) &\\
&  q$_{10}$  & 55 & (24,32,32,10,4,3,4) &\\
&  q$_3$,q$_6$ & 10,27 & (10,30,30,10,4,3,4) & \\
\hline
\end{tabular}
\caption{Basis used for the MCTDH simulations starting with a localised wavepacket on the Rydberg and zwitterionic states. The primitive basis was the harmonic oscillator DVR for every mode except the torsion, which uses a periodic FFT grid.  $N_i$ is the number of primitive basis per dimension in every particle, while $n_{i}$ is the number of SPFs used per electronic state. Config is the total number of configurations in the MCTDH wavefunction.} 
\label{basis-qua2}
\end{table}

\section{Conclusions}
We set out to investigate the importance of superpositions due to symmetrically equivalent minima to describe the initial wavepacket in photo-excited quantum dynamics simulations. The answer is found to be in the negative and no significant difference was seen in the excited state quantum dynamics of the 1,1-difluoroethylene molecular system using either localised (single minimum) or delocalised (symmetry-adapted) initial conditions. We observe that a qualitative correct behaviour of the deactivation of the V state is achieved with five degrees of freedom from the total of twelve. The torsion, CC and CF symmetric stretches, combination of asymmetric CF$_2$ stretch and CH$_2$ bend, and the CF$_2$ pyramidalisation drive the excited state dynamics on this system and explain the peaks observed in the absorption spectra after Fourier transforming the autocorrelation function. 

In contrast to previous SH dynamics,\cite{gomez2019} 
the current simulations find that there is barely population transfer to the electronic ground state, flowing into the Rydberg $\pi$-3s instead. This outcome is consistent for all three initial conditions investigated, illustrating that using a localised nuclear wavefunction is perfectly reasonable and the relative phases on the parts of the symmetric and antisymmetric superpositions do not affect the dynamics, leading to numerically identical results. There are minor differences for the full-dimensional system when starting with a localised wavefunction which maybe is due to convergence. As a way of validating our parameterised Hamiltonian, we performed dynamics using a localised initial condition on the valence, zwitterionic, and Rydberg electronic states and calculated absorption spectra weighted by the oscillator strengths that couple those states with the electronic ground state. An excellent agreement with the experimental spectrum measured in Reference~\citenum{limao2006} supports our model.

\section{Competing interests}
There are no competing interests to declare.

\section{Data availability}
Extra information such as ground state normal mode frequencies, potential energy curves along the twelve degrees of freedom and ab initio points against which the curves are fitted, parameters for the one-dimensional diabatic potentials and intra-state, linear and bilinear coupling parameters can be found in the Supporting Information. The quantics inputs and operator file are deposited at the UCL Repository https://rdr.ucl.ac.uk DOI: 10.5522/04/21318339.

\begin{acknowledgement}
We thank the MOLecules In Motion COST Action (CM1405) for funding short term scientific missions from the University of Vienna to University College London that made  this collaboration possible. SG acknowledges NextGenerationEU funds (Mar\'ia Zambrano Grant for the attraction of international talent). GW thanks the EPSRC for funding through grant EP/S028781/1.
\end{acknowledgement}

\bibliography{bibexport2}

\end{document}